\documentstyle[12pt]{article}
\begin{document}
\bibliographystyle{unsrt}
\vbox {\vspace{6mm}}
\begin{center}
{\large \bf DO THE ROBERTSON-SCHR\"{O}DINGER\\
 AND THE HEISENBERG\\
 UNCERTAINTY RELATIONS \\
 IMPLY A GENERAL PHYSICAL PRINCIPLE ?}\\[15mm]
{\bf Vinh Quang N.}\\[4mm]
{\it Institute of Physics, P.O.Box 429 Boho, Hanoi 10000, Vietnam}\\[10mm]
\end{center}
\vspace{2mm}
\begin{abstract}
It is explicitly shown that there exist physical states (normalized to 1) in which the Robertson- Schr\"{o}dinger and Heisenberg uncertainty relations are invalid, namely, the mean values of the physical operators are infinite. Consequently, these relations cannot imply a general physical principle.  The explanation by the theory of functional analysis is given : for these states even the definition of the uncertainty notion through the dispersion notion  in the probability theory is irrelevant
\vspace{0.5cm}
\end{abstract}
\section{Introduction}

      Since 1928 up to now, following Ruark [1], who, for the first time, called the Heisenberg uncertainty relation ( HUR) a principle, HUR and it's generalization, the Robertson- Schr\"{o}dinger uncertainty relation ( RSUR), are commonly looked upon as the mathematical representatives of a general physical principle. This implies that these relations must be valid for all physical states. After a modern, resent " which-way " experiment in an atom interferometer of Durr, Nonn and Rempe in 1998 [2] the discussion on the universality of the uncertainty principle  is exploded again. Therefore,  it is very useful to carefully reconsider the historical derivations of HUR and RSUR.\\

      In this paper, it is presented the remarks, notes and conclusions after careful reconsidering all of the derivations of (i) Heisenberg equality ( Heisenberg 1927 [3], (ii) Pauli inequality ( Kennard 1927 [4], Weyl 1928 [5]), (iii) Heisenberg uncertainty relation for two hermitian operators ( Robertson 1929 [6] ) and, (iv) Robertson-Schr\"{o}dinger uncertainty relation ( Robertson 1930 [7], Schr\"{o}dinger 1930 [8] ).
\vspace{1cm}
\section{Theory}

  a) First, the following remarks must be emphasized:\\
  
1. HUR does not forbid two physical quantities, whose corresponding operators are even noncommutative, to have simultaneously exact eigenvalues. This is due to Condon [9] who remarked that in the Hydrogen atom states with orbital quantum number l=0 and azimuthal quantum number m=0, the x-, y- and z-projections of the orbital angular momentum may all be exact zeros at the same time, although their operators do not commute with each other.\\

  2. Because of the speciality that their operators do not have any eigenfunction in  L$_2$
( space of the square integrable functions), each of both position and momentum has no exact value, which is also the direct  concequence of HUR . Hence, it is naturally redundant the statement that  position and momentum cannot simultaneously have exact values, which often is looked upon as the uncertainty principle.\\[5mm]
     b)   Second, we note that all the derivations implicitly contain superfine suppositions such as:\\

 1. The physical operators are always hermitian.\\

 2. In any normalized ( to 1) state the mean values of any physical operator are finite.\\

 That is, namely, the necessary condition for the validity of the definition of the notion "uncertainty",  introduced, for the first time, by Heisenberg for position and momentum in the Gauss wave packet case ( as a distance from the probability distribution maximum value to a value, at which the probability distribution decreases e times) and extended later by Weyl through the notion "dispersion" in the probability theory.\\[5mm]
 c)   Third, we point out some counter-examples which transparently show that 1. the physical operators are not always hermitian 2. the normalized state at which the mean value of the position operator is infinite and 3. the physical operators,  whose mean values in the ground state of Hydrogen atom are infinite.  They are following.\\

 1. Consider a eigen-differential wave packet of a 1-D free particle. It is normalized to unity. Let it is acted by the operator " position  cubed ". In acting on the obtained function the momentum operator is not hermitian.\\

 2. Consider an eigen-differential wave packet of a 1-D electron in a constant homogeneous electric field. It is normalized to unity. At this state the mean value of the position operator is infinite. Hence,  the uncertainty of the position cannot be defined. Then the statement (often called) of the uncertainty principle that the product of the position and momentum uncertainties much less than the certain (proportional to Planck ) constant, is meaningless.\\ 

 3. At the ground state of Hydrogen atom, the momentum operators to the n-th power with n equal or greater 6, for example, have infinite mean values, therefore, their uncertainties cannot be defined .\\

    While both the definition of the  uncertainty notion and the derivations are  valid for not-any physical state, RSUR and HUR cannot be recognized as a general physical principle; they might only be regarded as a " directe anschauliche Erlauterung" (direct intuitive interpretation) of the commutative relation between the corresponding operators, which was the original Heisenberg's point of view for the basic case of position and momentum [3].\\ 

    While RSUR and HUR do not imply the principle, it is not surprised that the obtained results in the " which-way" experiment  of Durr, Nonn and Rempe have no relation with RSUR and HUR.\\
   The above mentioned facts are explained by the theory of functional analysis as the following.\\

 1. Since both position and momentum operators are unbounded, they can get out of L$_2$ it's vectors. Then the physical operators are not sure of being hermitian.\\  

 2. Since 1926, when Hilbert, in answering to a Heisenberg's call, started investigating the mathematical foundations for quantum physics, after the half-century development of the functional analysis, the mathematicians obtained one of the most important results that is the following. The commonly defined expressions of  mean value and dispersion of the hermitian operators are meaningful only with the condition that the mean value is finite. But, at vectors outside of Schwartz subspace of the Hilbert space the mean values are not sure of being finite [10].
\section{Conclusion}

 1.RSUR and HUR are only two of the consequences of the current postulate ( axiom) systems in  quantum mechanics, whose the foundation is the statistical-probability description. Namely, RSUR and HUR are consequences of the hermitian property of the physical operators and normalizability (to 1) of the wave function. RSUR and HUR cannot be recognized as a general physical principle because they are only meaningful for not whole physical states.\\[5mm]
2. It is urgently necessary to make the following.\\

    (i) To change the concept for the uncertainty notion and, consequently, the notion of the certain value, exactly measured value.\\

    (ii) To improve the mathematical apparatus for quantum physics, namely, to use the rigged Hilbert space 
( Gelfand triplet) taking into account the fact that both of the momentum and position operators are unbounded, the importance of which is emphasized by mathematician-physicists such as Mackey, Fadeev, Maslov, Bogoliubov, Logunov and Todorov etc...\\

   (iii) To predict the new physical effects  using the newly improved mathematics.\\[1cm] 
{\bf ACKNOWLEGEMENT}\\

   The author would like to thank Profs. Y. S. Kim,  Chu Hao, Nguyen Van Hieu and Nguyen Van Dao 
for their interest in the present research. This work is supported in part by Vietnamese basic scientific research program 4.1.1601.


\newpage
\begin{thebibliography}{10}
\bibitem{1} Ruark A. E.: Phys. Rev. 1928, v. 31, p. 311
\bibitem{2} Durr S., Nonn T. and Rempe G.: Nature  1998, 395, 3 Sept.
\bibitem{3} Heisenberg W.: Z. Phys., 1927, Bd. 43, S. 172-198
\bibitem{4} Kennard E.H.: Z. Phys., 1927, Bd. 44, S. 326
\bibitem{5} Weyl J.: The Theory of Groups and Quantum Mechanics- N.Y.: Dover,1950.
\bibitem{6} Robertson H.P.: Phys. Rev. 1929, Vol. 34, p.163
\bibitem{7} Robertson H.P.: Phys. Rev. A. 1930, Vol. 35, P. 667
\bibitem{8} Schr\"{o}dinger E.: Berliner Berichte, 1930, S. 296
\bibitem{9} Condon E.U.: Science, 1929, v. 69, p.573
\bibitem{10} Functional analysis, ed. by G. Krein, Moscow 1972 ( in Russian )
\end{thebibliography}
\end{document}